\documentclass[twocolumn]{aastex631}
\usepackage[T1]{fontenc}
\usepackage{diagbox} 
\usepackage{makecell}
\usepackage{placeins}
\usepackage{ae,aecompl}
\usepackage{amstext}
\usepackage{url}
\usepackage{booktabs} 
\usepackage{array} 
\usepackage[]{natbib,amsmath,amssymb,times,bm}
\bibpunct{(}{)}{;}{a}{}{,}
\usepackage{tabularx,amsmath,amssymb}
\usepackage{mathrsfs} 
\usepackage{graphicx,epsfig,color,latexsym}
\usepackage{array}
\usepackage{booktabs}
\usepackage{multirow}
\usepackage{xcolor}
\usepackage{amssymb}
\usepackage{amsfonts}
\usepackage{xcolor}
\usepackage{hyperref}

\renewcommand{\d}{\mathrm{d}}
\newcommand{\e}{\mathrm{e}}

\newcommand{\bea}{\begin{eqnarray}}
\newcommand{\eea}{\end{eqnarray}}
\newcommand{\be}{\begin{equation}}
\newcommand{\ee}{\end{equation}}

\renewcommand{\exp}{\mathrm{exp}}

\begin{document}

\title{Predictions on the abundance of primordial black holes: results from the SMILE VLBI milli-lensing sample}

\correspondingauthor{Shuo Cao}
\email{caoshuo@bnu.edu.cn}

\author{Xinyue Jiang}
\affiliation{Institute for Frontiers in Astronomy and Astrophysics, Beijing Normal University, Beijing 102206, China;}
\affiliation{School of Physics and Astronomy, Beijing Normal University, Beijing 100875, China;}
\affiliation{National Centre for Nuclear Research, Pasteura 7, 02-093 Warsaw, Poland}

\author{Shuo Cao$^{\ast}$}
\affiliation{Institute for Frontiers in Astronomy and Astrophysics, Beijing Normal University, Beijing 102206, China;}
\affiliation{School of Physics and Astronomy, Beijing Normal University, Beijing 100875, China;}

\author{Marek Biesiada}
\affiliation{National Centre for Nuclear Research, Pasteura 7, 02-093 Warsaw, Poland}

\author{Xuanni Chen}
\affiliation{School of Physics and Astronomy, Beijing Normal University, Beijing 100875, China;}

\author{Yalong Nan}
\affiliation{Institute for Frontiers in Astronomy and Astrophysics, Beijing Normal University, Beijing 102206, China;}
\affiliation{School of Physics and Astronomy, Beijing Normal University, Beijing 100875, China;}

\author{Xinyang Jin}
\affiliation{Institute for Frontiers in Astronomy and Astrophysics, Beijing Normal University, Beijing 102206, China;}
\affiliation{School of Physics and Astronomy, Beijing Normal University, Beijing 100875, China;}

\author{Xuchen Tai}
\affiliation{Institute for Frontiers in Astronomy and Astrophysics, Beijing Normal University, Beijing 102206, China;}
\affiliation{School of Physics and Astronomy, Beijing Normal University, Beijing 100875, China;}

\begin{abstract}
Primordial black holes (PBHs) formed in the early universe represent a well-motivated dark matter candidate. However, their abundance in the intermediate-to-high mass range ($10^6$--$10^9\,M_\odot$) remains relatively less constrained. In this mass regime, the corresponding Einstein radii on the milliarcsecond scale make Very Long Baseline Interferometry (VLBI) the most direct observational probe via milli-lensing of compact radio sources. In this paper, we present predictions for the abundance of PBH dark matter ($f_{\rm PBH}$) using the large SMILE sample of $\sim 5000$ flat-spectrum compact radio sources. Under the optimistic observational conditions (with an angular resolution of $1.5\,\mathrm{mas}$, a maximum image separation of $100\,\mathrm{mas}$, and a maximum detectable flux ratio of $40$), assuming that no milli-lensing events are confirmed, we obtain the projected 95\% upper limit of $f_{\rm PBH}\lesssim0.12\%$ in the intermediate-to-high mass range, which would be approximately an order of magnitude tighter than previous VLBI milli-lensing limits. Broader lognormal mass functions yield correspondingly weaker projected limits. Generalising the optical depth to account for magnification bias and finite-core resolvability changes these projected limits by only about $5$\%. Moreover, we show that the predicted limits are highly sensitive to a small number of confirmed events, i.e., even one confirmed milli-lensing event would relax the projected upper limit to $0.20$\%, highlighting the critical importance of robust candidate confirmation in the SMILE sample. Finally, under the optimistic baryonic accretion scenario, the PBH mass growth changes the projected limits by $\lesssim 1$\% within the VLBI-sensitive mass range, confirming the robustness of our results.
\end{abstract}


\section{Introduction} \label{sec:intro}

The physical nature of dark matter remains one of the most important open questions in modern cosmology and astrophysics. Observations from galaxy rotation curves, galaxy cluster dynamics, the cosmic microwave background (CMB), and large-scale structure formation all indicate that most of the matter in the Universe exists in the form of non-baryonic dark matter \cite[][]{1980ApJ...238..471R,2020A&A...641A...6P}. However, its composition has not yet been directly detected. In addition to particle dark matter candidates, primordial black holes (PBHs), formed in the early universe, have long been considered as a possible component that could account for part or even all of the dark matter \cite[][]{1974MNRAS.168..399C,2016PhRvD..94h3504C}. If the amplitude of density perturbations in a given region is sufficiently large, the region can collapse directly into a black hole under gravity when it re-enters the horizon. Thus, one key feature of PBHs is that their masses can span a very wide range, mainly depending on the horizon scale at formation and the specific formation mechanism, which in turn encodes information about small-scale primordial density fluctuations. During the radiation-dominated era, enhanced small-scale density fluctuations after inflation, cosmological phase transitions, or the collapse of topological defects can all lead to PBH formation over a broad range of masses \cite[][]{2010PhRvD..81j4019C,2021JPhG...48d3001G}. In general, the initial mass of PBHs is closely related to the horizon mass at formation and can range from sub-stellar scales to intermediate-mass black holes or even higher. After formation, PBHs may further evolve in mass through accretion. Baryonic accretion can lead to mass growth, but this process strongly depends on the environment, such as gas density, relative velocity, and the geometry of the accretion flow \cite[][]{1952MNRAS.112..195B,2008ApJ...680..829R,2017PhRvD..95d3534A}. As a result, PBHs in different environments may follow very different evolutionary paths: those residing in dense dark matter halos may grow significantly, while those in low-density cosmic backgrounds typically retain nearly constant mass. Therefore, whether PBHs preserve their initial masses or undergo substantial growth is crucial for interpreting observational constraints on their abundance. In recent years, observations from JWST of massive galaxies at high redshift ($z \gtrsim 7$) have triggered extensive discussions on early black hole formation \cite[][]{2022ApJ...940L..14N,2022ApJ...938L..15C,2024SCPMA..6789511T}. One possible explanation is that the central black holes in these galaxies originate from PBHs with initial masses of $\sim 10^3M_\odot$, which subsequently grow rapidly to $10^6$--$10^8M_\odot$ through efficient or even super-Eddington accretion in dense environments \cite[][]{2025JCAP...04..040Z,2024SCPMA..6789511T}. This scenario points to a particularly relevant PBH mass range around $10^6$--$10^8M_\odot$, which is of interest both for early structure formation and for current observational constraints. Many studies have shown that in the low-redshift Universe ($z \lesssim 1$), the accretion efficiency of PBHs drops significantly due to the decrease in baryon density caused by cosmic expansion \cite[][]{2008ApJ...680..829R,2017PhRvD..95d3534A}. Since PBH masses at low redshift largely reflect their formation values, observational constraints on their present-day abundance and mass distribution directly probe the initial mass function and thus serve as indirect probes of the small-scale primordial power spectrum, complementing large-scale probes such as the CMB.

Constraints on the abundance of primordial black holes (PBHs) rely on different astrophysical probes across different mass ranges. At the low-mass end, gravitational microlensing surveys constrain PBHs in the range $10^{-9}-10^{4} M_\odot$ \citep{2025JCAP...04..023G}, while gravitational wave observations from binary mergers probe $10^{-6}-10^{2} M_\odot$ \citep{2024CQGra..41n3001D}, and fast radio burst (FRB) lensing provides complementary sensitivity around the stellar-mass window \citep{2022MNRAS.511.1141Z}. At higher masses, CMB constraints arise from two distinct mechanisms: energy injection due to PBH accretion during the cosmic dark ages, which constrains masses $\sim 10 M_\odot$ up to $\sim 10^4 M_\odot$ \citep{2020PhRvR...2b3204S,2017PhRvD..96h3524P}, and $\mu$-type spectral distortions sourced by the dissipation of density perturbations associated with PBH formation, which constrain the range $10^5-10^{12} M_\odot$ \citep{2018PhRvD..97d3525N}. However, while the $\mu$-distortion bound formally overlaps with our mass range of interest, it applies specifically to PBHs formed directly from primordial inhomogeneities and is therefore sensitive to assumptions about the formation mechanism. In contrast, gravitational lensing provides a purely geometric and model-independent probe of compact objects regardless of their formation history. In the mass range $10^6-10^9 M_\odot$, the corresponding Einstein radii on the milliarcsecond scale make Very Long Baseline Interferometry (VLBI) the most direct observational probe via milli-lensing of compact radio sources. Several previous works have used this approach to constrain the abundance of massive compact objects in this range. For example, \citet{2001PhRvL..86..584W} searched for milli-lens candidates with angular separations of $1.5$-$50$ mas using multi-frequency VLBI observations of $\sim 300$ sources and derived an upper limit of $\Omega_{\rm CO} < 0.01$ for compact objects with masses
between $10^6$ and $10^8$ solar masses, while later \citet{2022MNRAS.513.3627Z} performed a selection based on \cite{1985AJ.....90.1599P} and \cite{2006JCAP...11..002J} and used several hundred flat-spectrum VLBI compact radio sources to explore larger separation ranges, and additionally applied the same formalism to the Astrogeo-based SMILE pilot sample of \cite{2021MNRAS.507L...6C}. These analyses share a feature that a larger sample directly leads to stronger statistical upper limits. With the rapid development of VLBI databases and radio surveys, the number of compact radio sources with high-resolution images has increased significantly, making it possible to improve these constraints. The recently constructed SMILE (Search for MIlli-LEnses) project provides a large VLBI dataset built along this direction, combining the CLASS survey with VLBI imaging to provide a sample of several thousand compact radio sources \cite[][]{2021MNRAS.507L...6C,2022A&A...668A.166L}. Compared to earlier samples with only a few hundred sources, SMILE offers a substantial improvement in sample size, and a
large fraction of the sources have measured redshifts, allowing the source redshift distribution to be explicitly incorporated into lensing probability calculations.

In this paper, we use the preliminary estimated size of the SMILE VLBI data sample and the redshift distribution of sources, as presented in \citet{2022A&A...668A.166L} to provide predictions on PBH abundance from milli-lensing. Specifically, we make the following advances beyond previous works: (i) we consider the SMILE sample of 4968 flat-spectrum compact radio sources---nearly an order of magnitude larger than previous analyses, and perform scenario-based estimate quantifying how sensitive the predicted limits are to the confirmation of even a single milli-lensing event; (ii) we examine both monochromatic and lognormal PBH mass functions, quantifying how an extended mass distribution degrades the constraining power; (iii) we generalize the optical-depth calculation to account for magnification bias and finite-core resolvability, which reduces to the purely geometric treatment adopted in previous analyses in the fiducial limit. These corrections, though small in the current sample, may become important when extending this analysis to surveys with different frequencies, resolutions, or source populations; and (iv) we investigate the impact of baryonic accretion on PBH mass evolution and show that the resulting projected limits remain robust even under optimistic accretion assumptions. Together, these set out the projected reach of a milli-lensing search on $f_{\rm PBH}$ in the mass range of $10^6\text{--}10^9\,M_\odot$. Throughout the analysis, we adopt a flat $\Lambda$CDM cosmology with $H_0 = 70\ \mathrm{km\ s^{-1}\ Mpc^{-1}}$, $\Omega_m = 0.3$, and $\Omega_{\mathrm{DM}} = 0.26$.


\section{Methodology}
\label{sec:Theoretical Framework}

To predict the abundance of PBHs using VLBI milli-lensing observations, we compute the probability that a background radio source is detectably lensed under the observational constraints of VLBI imaging. The key idea is to map the observational selection criteria into an allowed region in the source plane, which then defines an effective lensing cross section and the corresponding optical depth \cite[][]{1973ApJ...185..397P}. Since PBHs are extremely compact compared with the relevant lensing scales, they can be treated as point-mass lenses. For a lens of mass $M$ at redshift $z_L$ and a source at redshift $z_S$, the Einstein radius is given by \citep{2012JCAP...03..016C,2015ApJ...806..185C}
\begin{equation}
\theta_E(M,z_L,z_S)=
\sqrt{\frac{4GM}{c^2}\frac{D_{LS}}{D_LD_S}},
\end{equation}
where $D_L$, $D_S$, and $D_{LS}$ are the angular diameter distances from the observer to the lens, from the observer to the source, and from the lens to the source, respectively.

For a point-mass lens, all observable lensing quantities are determined by the dimensionless impact parameter $y \equiv \beta/\theta_E$, where $\beta$ is the angular position of the source relative to the lens. Whether a milli-lensing event can be identified in VLBI images depends on both the angular separation of the two images ($\Delta\theta$) and their flux ratio ($R_f$): the separation must be resolved, while the flux ratio must be small enough for the fainter image to be detected within the finite dynamic range of the observation. For a point-mass lens, both observables can be written as functions of the impact parameter ($y$) 

\begin{equation}
\begin{aligned}
\Delta\theta(y) &= \theta_E\sqrt{y^2+4}, \\[2pt]
R_f(y) &= \frac{y^2+2+y\sqrt{y^2+4}}{\,y^2+2-y\sqrt{y^2+4}\,}.
\end{aligned}
\end{equation}
The image separation must satisfy $\delta_{\rm eff} < \Delta\theta(y) < \Delta$, where the lower limit is set by the effective angular resolution $\delta_{\rm eff}$ and the upper limit $\Delta$ is the largest image separation searched in the observational window. Since $R_f(y)$ increases monotonically with $y$, requiring the flux ratio to stay within the dynamic range, $R_f(y)\le R_{f,\max}$, is equivalent to an upper bound on the impact parameter, $y\le y_R\equiv R_{f,\max}^{1/4}-R_{f,\max}^{-1/4}$.
Combining these conditions, the resolution and window limits give
\begin{equation}
\begin{aligned}
y_{\min}(M,z_L,z_S) &= \sqrt{\max\!\left[(\delta_{\rm eff}/\theta_E)^2-4,\;0\right]}, \\[2pt]
y_{\max}(M,z_L,z_S) &= \min\!\left[\,\sqrt{\max\!\left[(\Delta/\theta_E)^2-4,\;0\right]}\,,\; y_R\,\right].
\end{aligned}
\label{eq:ybounds}
\end{equation}
For a given $(z_L,z_S)$, not all lens masses produce detectable milli-lensing configurations. At the high-mass end, the minimum image separation of a point-mass lens is $2\theta_E$; once $2\theta_E>\Delta$, no impact parameter satisfies $\Delta\theta(y)<\Delta$ and the allowed region closes. At the low-mass end, $\theta_E$ is small, so the separation $\delta_{\rm eff}$ can only be reached at large $y$; but a larger $y$ also gives a larger flux ratio, so $y_{\min}$ eventually exceeds $y_R$ and the fainter image falls below the dynamic-range limit. The allowed region therefore exists only when $y_{\max}>y_{\min}$. Near both boundaries the allowed range of impact parameters narrows progressively, the effective lensing cross section decreases smoothly, and the sensitivity weakens towards the edges of the testable mass range.

For an ideal point source the resolvability of the two images is set by the nominal VLBI angular resolution $\delta$. In milli-lensing, however, the image separation is of order milliarcseconds and can be comparable to the apparent core size $\theta_{\rm core}$ of a compact radio source, so the finite source size can affect whether the two images are reliably deblended. We therefore combine the instrumental resolution and the source-size limitation into an effective resolution $\delta_{\rm eff}=\max(\delta,\kappa\,\theta_{\rm core})$, where we adopt $\kappa=2$, i.e. two images are taken to be separable only when their separation exceeds twice the core size. In the point-source limit $\theta_{\rm core}\to0$ this reduces to $\delta_{\rm eff}=\delta$.
A flux-limited source sample is also subject to magnification bias. Because lensing increases the observed flux of a background source, intrinsically fainter sources that would otherwise fall below the survey flux limit can be magnified into the sample. As a result, the detection probability is weighted not only by the source-plane area but also by the magnification. Since the sample is selected on the total flux density of the unresolved source, the relevant quantity is the total magnification, which for a point-mass lens is
\begin{equation}
\mu_{\rm tot}(y) = \frac{y^2+2}{y\sqrt{y^2+4}}.
\end{equation}
For a radio sample whose cumulative counts above a flux density $S$ follow $N(>S)\propto S^{-\eta}$, a magnification $\mu_{\rm tot}$ lowers the effective flux limit from $S$ to $S/\mu_{\rm tot}$ and raises the number of sources entering the sample by a factor $\mu_{\rm tot}^{\eta}$; at the same time it stretches the source plane and decreases the source surface density by $\mu_{\rm tot}^{-1}$ \cite[][]{1992ARA&A..30..311B}. The magnification-bias weight is therefore $\mu_{\rm tot}^{\eta}\times\mu_{\rm tot}^{-1}=\mu_{\rm tot}^{\eta-1}$, which reduces to unity for $\eta=1$, when the sources gained from the lowered flux limit are exactly cancelled by the area dilution.

The above mapping shows that the VLBI selection function effectively defines an annular region of allowed impact parameters, $y_{\min}<y<y_{\max}$. The corresponding lensing cross section in the lens plane is therefore given by

\begin{equation}
\sigma_{\rm lens}(M,z_L,z_S)=\pi\,(D_L\theta_E)^2\,\mathcal{K}(M,z_L,z_S)
\end{equation}
where the dimensionless detectability kernel is
\begin{equation}
\mathcal{K}(M,z_L,z_S)=\int_{y_{\min}}^{y_{\max}} 2y\,\mu_{\rm tot}(y)^{\eta-1}\,dy.
\end{equation}

The mass integral formally extends over all masses, but $\mathcal{K}$ vanishes wherever $y_{\max}\le y_{\min}$, so only the mass range in which the VLBI thresholds admit a solution contributes.
Assuming that PBHs constitute a fraction $f_{\rm PBH}$ of dark matter, the probability that a source at redshift $z_S$ is detectably lensed is described by the optical depth \cite[][]{2012ApJ...755...31C}, which can be written as
\begingroup
\thinmuskip=1mu
\medmuskip=2mu
\thickmuskip=3mu
\begin{equation}
\hspace*{-0.35cm}
\begin{aligned}
&\tau(z_S,M_{\rm PBH},f_{\rm PBH})
= \int_0^{z_S} d\chi \, (1+z_L)^2 \int_0^{\infty} dM \, \sigma_{\rm lens} \, n(M,z_L) \\[6pt]
&= \int_0^{z_S} \frac{c \, dz_L}{H(z_L)} (1+z_L)^2 \int_0^{\infty} dM \, \pi \theta_E^2(M,z_L,z_S) D_L^2 \\
&\quad \times
\mathcal{K}(M,z_L,z_S)
\cdot \frac{f_{\rm PBH} \Omega_{\rm DM} \rho_{c,0} \psi(M)}{M} \\[6pt]
&= \frac{3}{2} f_{\rm PBH} \Omega_{\rm DM} \int_0^{z_S} dz_L \, \frac{H_0^2}{c H(z_L)} \frac{D_L D_{LS}}{D_S} (1+z_L)^2 \\
&\quad \times \int_0^{\infty} dM \, \psi(M)\,
\mathcal{K}(M,z_L,z_S)
\end{aligned}
\label{eq:tau}
\end{equation}
\endgroup
where $\Omega_{\rm DM}$ is the present-day dark matter density parameter, $n(M,z_L)$ is the comoving number density of PBHs of mass $M$ and $H(z)$ is the Hubble parameter. Throughout, $M$ denotes the mass of an individual lens and is integrated over, while $M_{\rm PBH}$ labels the mass function itself; for the monochromatic case the two coincide. $\psi(M)$ denotes the normalized PBH mass distribution function, which describes the relative abundance of PBHs across different masses and satisfies the normalization condition $\int \psi(M) \mathrm{d}M = 1$. Since $\mathcal{K}=0$ for $y_{\max}\le y_{\min}$ (Eq.~\ref{eq:ybounds}), only $M\in[M_{\min}(z_L,z_S),M_{\max}(z_L,z_S)]$ contributes to the integral. For $\eta=1$ and $\theta_{\rm core}\to0$ the weight is unity and $\mathcal{K}$ reduces to the purely geometric area $y_{\max}^2-y_{\min}^2$ adopted in previous VLBI milli-lensing analyses \citep{2022MNRAS.513.3627Z}, which is thus recovered as a special case of the present framework.


In a realistic VLBI sample, the sources are distributed over a range of redshifts rather than located at a single $z_S$. The single-source optical depth is averaged over the normalized source redshift distribution $N(z_S)$, adopting the source redshift distribution of the SMILE project presented in \citet{2022A&A...668A.166L},
\begin{equation}
\bar{\tau}(M_{\rm PBH},f_{\rm PBH})=\int {\rm d}z_S\,N(z_S)\,\tau(z_S;M_{\rm PBH},f_{\rm PBH}),
\end{equation}
where $\int {\rm d}z_S\,N(z_S)=1$. This quantity represents the optical depth of a single source, averaged over the source redshift distribution. The total expected number of detectable milli-lensing events over the whole sample of $N_{\rm src}$ sources follows as
\begin{equation}
N_{\rm lens}=N_{\rm src}\left(1-e^{-\bar{\tau}}\right),
\end{equation}
Since $\bar{\tau}\ll 1$ is typically satisfied in the milli-lensing regime, we have $N_{\rm lens}\simeq N_{\rm src}\bar{\tau}$ as a good approximation.

In the absence of confirmed milli-lensing detections, the number of observed events is zero, and the Poisson probability of this outcome yields a 95\% confidence upper limit on the PBH abundance through $N_{\rm src}\bar{\tau}=2.996,$
\begin{equation}
f_{\rm PBH}^{95\%}(M_{\rm PBH})
= \frac{2.996}{N_{\rm src}\,\int {\rm d}z_S\,N(z_S)\,
\tau(z_S;M_{\rm PBH},f_{\rm PBH}=1)}.
\label{eq:fpbh95}
\end{equation}


From the above equations, it is clear that the theoretical optical depth is highly sensitive to the $M_{\rm PBH}$. Here we first consider a monochromatic PBH mass function
\begin{equation}
\psi(M)=\delta(M-M_{\rm PBH}),
\end{equation}
in which all PBHs have the same mass. In this case, the predictions on $f_{\rm PBH}$ at a given mass should be interpreted as the maximum allowed dark matter fraction under this assumption.

More generally, PBH formation scenarios often predict extended mass distributions \cite[][]{2016PhRvD..94h3504C,2021JPhG...48d3001G}. To account for this, we also consider a lognormal mass function
\begin{equation}
\psi(M)=\frac{1}{\sqrt{2\pi}\sigma M}
\exp\left[-\frac{\ln^2(M/M_c)}{2\sigma^2}\right],
\label{eq:lognormal}
\end{equation}
where $M_c$ is the central mass and $\sigma$ characterizes the width of the distribution. The function $\psi(M)$ is normalized to unity, and the total PBH abundance is still described by a single parameter $f_{\rm PBH}$. For the lognormal mass function, substituting Eq.~(\ref{eq:lognormal}) into Eq.~(\ref{eq:tau}) gives

\begin{equation}
\begin{aligned}
&\tau(z_S; M_c, \sigma, f_{\rm PBH})
= \frac{3}{2} f_{\rm PBH}\Omega_{\rm DM}\int_0^{z_S} {\rm d}z_L\,
\frac{H_0^2}{cH(z_L)} \times\\
&\frac{D_LD_{LS}(1+z_L)^2}{D_S}
\int_0^{\infty}{\rm d}M\,
\frac{1}{\sqrt{2\pi}\sigma M}
\exp\left[-\frac{\ln^2(M/M_c)}{2\sigma^2}\right] \times\\
&\mathcal{K}(M,z_L,z_S).
\end{aligned}
\label{eq:tau_logn}
\end{equation}
In addition, if PBHs experience mass growth due to accretion, the lens mass entering the above expressions should be generalized to $M = M(M_0, z_L)$. The impact of such mass evolution on the predicted constraints will be discussed in Section {\ref{sec:results}}.

\section{Observational data} \label{sec:Data}

The statistical power of VLBI milli-lensing constraints is primarily determined by the number of compact radio sources and their redshift distribution, in addition to the instrumental resolution \citep{2015ApJ...806...66C}. To facilitate comparison with earlier VLBI milli-lensing studies, we first briefly introduce the sample from the 2.29 GHz VLBI survey \citep{1985AJ.....90.1599P}, which observed a large set of radio sources and identified 917 objects with compact structure above a correlated flux density limit of $\sim 0.1$ Jy. \cite{2006JCAP...11..002J} later compiled updated redshift and radio information for 613 of these sources, covering a redshift range of $0.0035 \le z \le 3.787$. Based on this catalog, \cite{2017JCAP...02..012C,2017A&A...606A..15C,2022MNRAS.513.3627Z} further selected flat-spectrum core-dominated sources with spectral index $\alpha \ge -0.5$, which are well suited for identifying milliarcsecond-scale multiple-image configurations. This yields a sample of 543 flat-spectrum compact radio sources (hereafter referred to as the Preston sample). This sample represents one of the earliest statistically defined VLBI datasets used for milli-lensing searches and provides a useful reference for comparison with more recent surveys.

Compared to earlier VLBI samples, the development of VLBI databases and radio surveys in recent years has significantly increased the number of compact radio sources available for milli-lensing searches. The main sample used in this work comes from the SMILE (Search for MIlli-LEnses) project \cite[]{2021MNRAS.507L...6C,2022A&A...668A.166L}, which constructed a large VLBI dataset. The SMILE sample is based on the CLASS (Cosmic Lens All-Sky Survey) catalog \cite[][]{2003MNRAS.341....1M}, which contains 11,685 flat-spectrum radio sources selected from the GB6 (5 GHz) and NVSS (1.4 GHz) surveys over declinations $0^\circ \le \delta \le 75^\circ$. The CLASS selection requires a 5 GHz flux density above 30 mJy, a flat spectrum between 1.4 and 5 GHz, and a Galactic latitude $|b| \ge 10^\circ$. Building on this, SMILE further selects sources with total flux density above 50 mJy at 8 GHz, resulting in a final sample of 4968 flat-spectrum compact radio sources. Optical counterparts are identified within $3''$ by cross-matching with the OCARS compilation of optical characteristics of VLBI sources, which provides heterogeneous redshift information for about two-thirds of the sample \cite[][]{2022A&A...668A.166L}. Thanks to its relatively uniform selection strategy inherited from CLASS, the selection function of the SMILE sample is comparatively well characterized, allowing a more controlled and interpretable lensing probability calculation. We therefore adopt this dataset as our primary sample (hereafter referred to as the SMILE sample). In this work, we use the redshift distribution $N(z_s)$ derived from the two-thirds of SMILE sources with measured redshifts as an estimate of the overall source redshift distribution, implicitly assuming that the known-redshift subsample is representative of the full sample \cite[][]{2025ApJS..276...38P, 2021MNRAS.507L...6C, 2025A&A...695A.169P}.

\section{Results and discussion} 
\label{sec:results}

The VLBI milli-lensing constraints on PBH abundance depend not only on the theoretical model, but also sensitively on the observational setup, including the angular resolution, the effective search field, and the range of detectable flux ratios. For the Preston sample, the $2.29$~GHz survey of \cite{1985AJ.....90.1599P} provides only the parent source list; the null milli-lensing result adopted for this sample comes from the $5$~GHz VLBI search of \cite{2001PhRvL..86..584W}, which reports no multiple images over the separation range $1.5$--$50$~mas. \cite{2006JCAP...11..002J} expanded the source counts to several hundred, then \cite{2022MNRAS.513.3627Z} additionally impose a flat-spectrum selection ($\alpha\ge-0.5$), yielding a final sample of 543 sources for the milli-lensing null-detection analysis. We therefore take $\delta = 1.5\,\mathrm{mas}$ as the angular resolution and set the maximum detectable image separation to $\Delta = 50\,\mathrm{mas}$ following their treatment. Here, $\delta$ represents the typical angular resolution of GHz VLBI observations, while $\Delta$ reflects the effective field of view used to search for multiple compact components in early VLBI imaging, corresponding to a characteristic scale of about $50 \times 50\,\mathrm{mas}$. For the SMILE sample based on the CLASS survey, we adopt $\delta = 1.5\,\mathrm{mas}$ as a representative angular-resolution threshold, consistent with the C-/X-band VLBA archival observations planned for the milli-lensing search \cite[]{2022A&A...668A.166L}. The SMILE project is designed to probe milli-lensing image separations on $\lesssim 100\,\mathrm{mas}$ \cite{2022A&A...668A.166L}. Motivated by this characteristic angular scale, we adopt $\Delta = 100\,\mathrm{mas}$ as the effective upper bound of the VLBI search window. This larger search window extends the sensitivity of the SMILE sample toward lenses with larger Einstein radii, and thus to higher PBH masses. For the flux ratio threshold, we adopt two representative values to describe different observational conditions. A commonly used reference in strong lensing studies is $R_{f,\max} = 7$, which traces back to the classical analysis of magnification bias in lensing statistics \citep{1984ApJ...284....1T}. On the other hand, high-resolution radio observations can probe significantly larger flux contrasts, following \citet{2001PhRvL..86..584W}, we take $R_{f,\max} = 40$ as an optimistic observational limit.

\begin{figure}
 \centering
 \includegraphics[width=\linewidth]{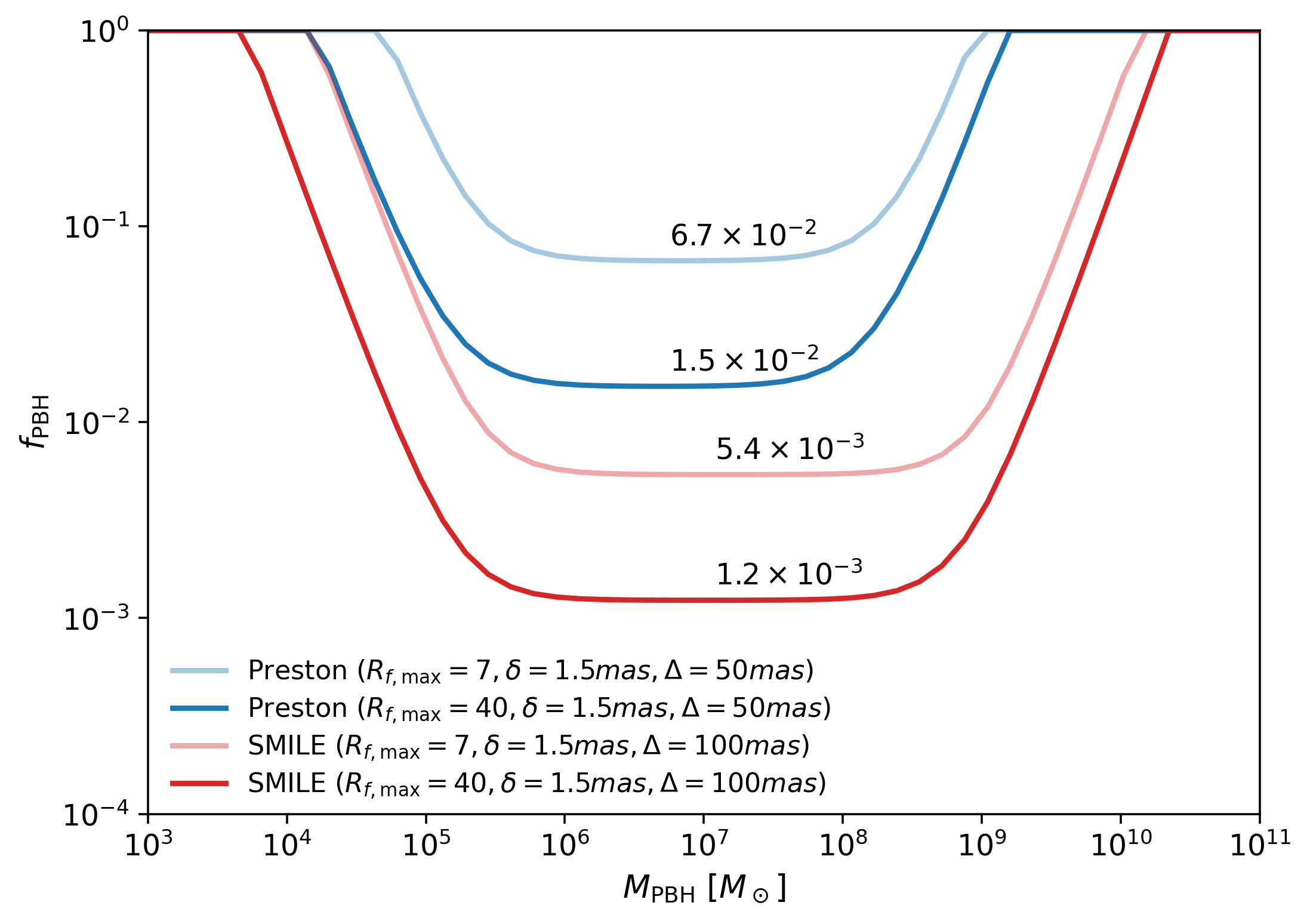}
 \caption{Predicted 95\% upper limits on the PBH dark matter fraction $f_{\rm PBH}$ as a function of PBH mass for the monochromatic mass function. Blue curves correspond to the Preston sample, while red curves correspond to the SMILE sample. Different color shades indicate different flux-ratio thresholds, with $R_{f,\max}=7$ (lighter colors) and $R_{f,\max}=40$ (darker colors). The observational parameters are $\delta=1.5\,\mathrm{mas}$ for all cases, with $\Delta=50\,\mathrm{mas}$ for Preston and $\Delta=100\,\mathrm{mas}$ for SMILE. The numerical labels indicate the minimum $f_{\rm PBH}$ values reached by each curve.}
 \label{fig:mono}
\end{figure}

\begin{figure}
 \includegraphics[width=\linewidth]{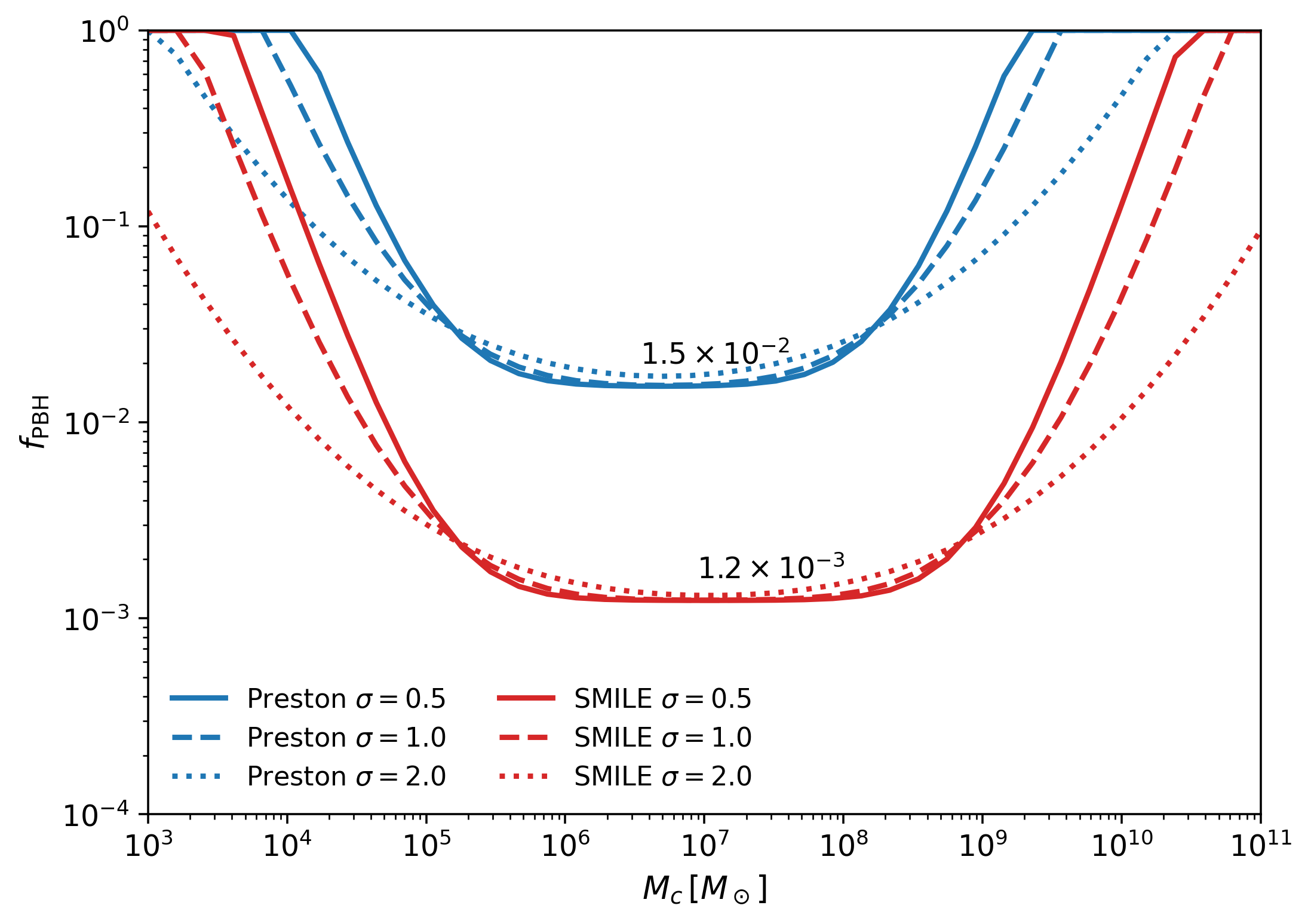}
 \caption{\textbf{The same as Fig.~\ref{fig:mono}. Here we consider a lognormal PBH mass function, with the x-axis showing the characteristic mass $M_c$.} We adopt $\delta=1.5\,\mathrm{mas},\,\Delta=50\,\mathrm{mas}$, $R_{f,\max}=40$ for the Preston sample and $\delta=1.5\,\mathrm{mas},\,\Delta=100\,\mathrm{mas}$, $R_{f,\max}=40$ for the SMILE sample. The solid, dashed, and dotted curves correspond to $\sigma=0.5,\ 1.0,\ 2.0$, respectively. The numerical labels indicate the minimum $f_{\rm PBH}$ for the case of $\sigma=0.5$.}
 \label{fig:log}
\end{figure}

The predictions for the monochromatic PBH case are shown in Fig.~\ref{fig:mono}. These predictions correspond to the fiducial, purely geometric detectability kernel $(\eta=1, \theta_{\rm core}\to0)$, for which $\mathcal{K}=y_{\max}^2-y_{\min}^2$ and our framework reduces to that adopted in previous VLBI milli-lensing analyses, enabling a direct comparison with the Preston sample. The magnification-bias and finite-core corrections are discussed further below. For the Preston sample we adopt the null search result of \citet{2022MNRAS.513.3627Z}, whereas the SMILE search is still ongoing, and we assume zero detections when deriving the projected limits. The curves from different samples and different flux ratio thresholds exhibit a similar shape, i.e., the strongest limits appear around an intermediate mass range, while they become weaker toward both the low-mass and high-mass ends. This behavior directly reflects the VLBI observational window. At low masses, the Einstein radius is too small, leading to image separations below the resolution limit $\delta$; at high masses, the image separation becomes too large and exceeds the upper bound $\Delta$. As a result, the effective lensing cross section peaks at intermediate masses. As expected from standard lensing statistics, for the same $R_{f,\max}$, the SMILE sample (red curves) yields systematically stronger projected limits than the Preston sample (blue curves), with lower $f_{\rm PBH}$ upper limits across the full mass range. In addition, due to the larger search window, the SMILE sample extends its reach toward higher masses. The choice of flux ratio threshold also has a clear impact: a larger $R_{f,\max}=40$ corresponds to a wider allowed range of impact parameter (i.e., larger $y_{\max}$), increasing the effective lensing cross section and thus would lead to stronger constraints, while a smaller threshold $R_{f,\max}=7$ gives more conservative and thus weaker limits. This figure is also consistent with the expected trend that increasing the sample size is the dominant factor in improving the future constraints: even comparing the Preston sample at $R_{f,\max}=40$ with the SMILE sample at the more conservative $R_{f,\max}=7$, the latter would still yield stronger constraints across the intermediate mass range, because the nearly order-of-magnitude increase in source count outweighs the gain from an enhanced flux-ratio sensitivity. We also consider a future SMILE scenario in which only one of the 4968 parent sources is confirmed as a milli-lensing event. In this case, $N_{\rm lens}$ changes from zero to one, the 95\% Poisson bound becomes $N_{\rm src} \bar{\tau} = 4.74$, leading to a less stringent upper limit on $f_{\rm PBH}$, from 0.12\% to 0.20\%. This shows that the result is sensitive to the number of confirmed lensing events, making candidate confirmation essential.

In PBH studies, the monochromatic mass function is usually treated as an idealized reference. If the formation mechanism corresponds to a primordial power spectrum with finite width, or if subsequent evolution broadens the initial mass distribution, then the PBH mass function is more naturally described by an extended distribution with a characteristic mass and finite width. For this reason, we extend the analysis using a lognormal mass function (see Eq.~\ref{eq:lognormal}), to examine how the probability weight is distributed relative to the mass range where VLBI milli-lensing is most sensitive. In this model, $M_c$ represents the characteristic mass, and $\sigma$ describes the width of the distribution in $\ln M$. In the limit $\sigma \rightarrow 0$, the lognormal distribution reduces to the monochromatic case. As $\sigma$ increases, the distribution becomes broader, spreading probability weight toward both lower and higher masses. Representative values of $\sigma$ ranging from relatively narrow to broad distributions have been widely adopted in the PBH literature \citep{2017JCAP...09..037R,2017PhRvD..96b3514C, 2019EPJC...79..717L}, and we adopt $\sigma = 0.5,\ 1.0,\ 2.0$ as three illustrative cases. In this part, we fix $R_{f,\max}=40$ to highlight the constraining power under optimistic observational conditions. \textbf{The results in Fig.~\ref{fig:log} show that as $\sigma$ increases from 0.5 to 2.0, the location of the minimum of the curves does not change significantly for either sample, while the overall amplitude increases and the curves broaden toward both low and high masses.} The comparison between the two samples follows the same trend as in the monochromatic case: for all values of $\sigma$, the SMILE sample would always yield stronger predicted limits than Preston's. Physically, a larger $\sigma$ corresponds to a more extended mass distribution, so a smaller fraction of PBHs falls within the mass window where VLBI milli-lensing is most sensitive, leading to systematically weaker predicted limits. This also implies that applying monochromatic constraints directly to extended mass functions may overestimate the constraining power, particularly when a significant fraction of the distribution lies outside the peak sensitivity window.

\begin{figure}
 \centering\includegraphics[width=\linewidth]{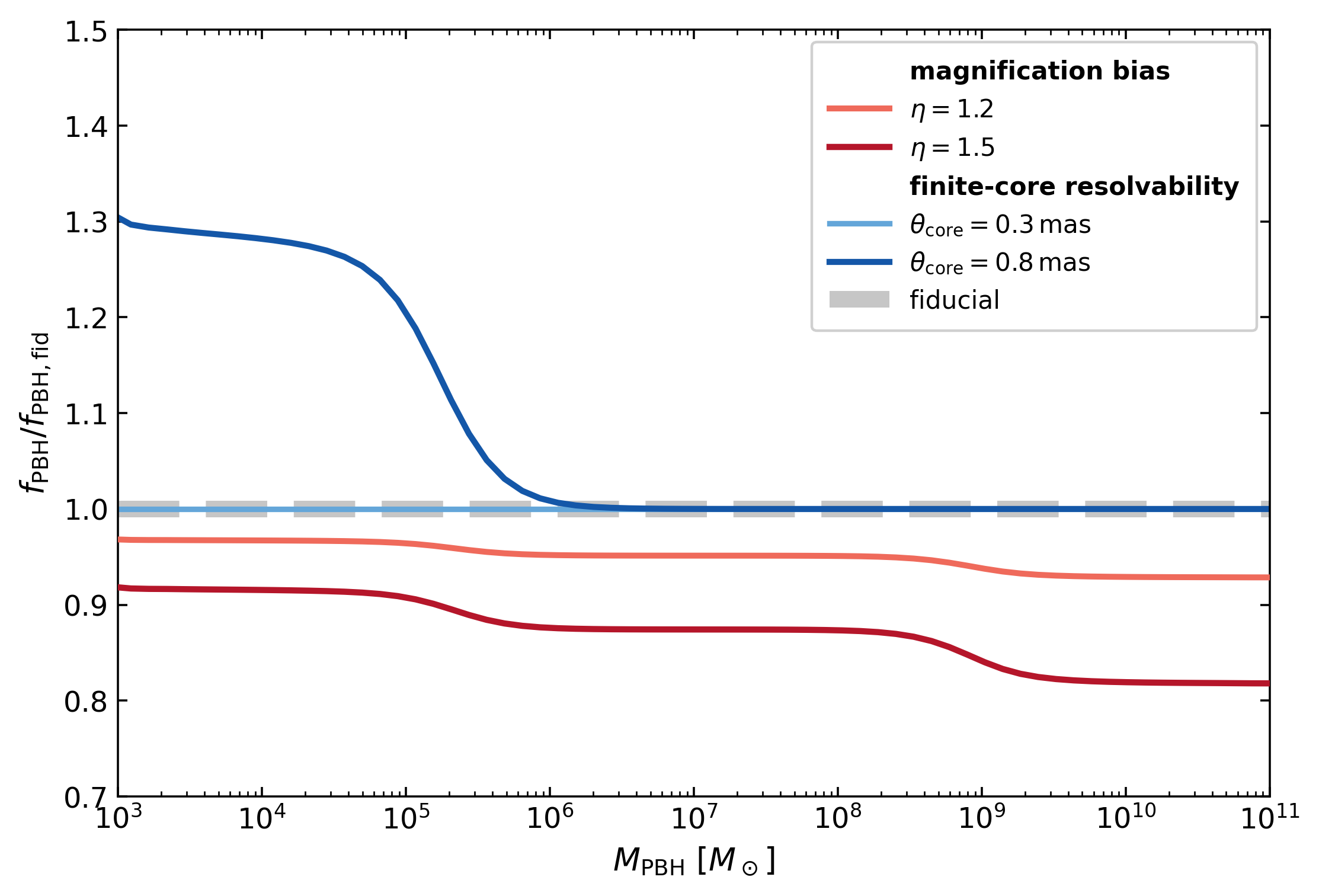}
 \caption{Ratio of the SMILE 95\% confidence upper limit on \(f_{\rm PBH}\) to the fiducial geometric result as a function of PBH mass. Red curves show magnification-bias corrections for cumulative source-count slopes \(\eta=1.2\) and \(1.5\), where \(N(>S)\propto S^{-\eta}\). Blue curves show finite-core corrections for \(\theta_{\rm core}=0.3\) and \(0.8\,\mathrm{mas}\). The thick light-grey dashed line denotes the fiducial case, \(\eta=1\) and \(\theta_{\rm core}=0\). All other parameters are fixed to the SMILE setup: \(N_{\rm src}=4968\), \(\delta=1.5\,\mathrm{mas}\), \(\Delta=100\,\mathrm{mas}\), and \(R_{f,\max}=40\).}
 \label{fig:corrections}
\end{figure}

We now examine how the predictions change once the magnification-bias and finite-core effects are included. For the magnification bias, which increases the effective lensing probability by boosting faint sources above the flux limit, each source-plane area element is weighted by $\mu_{\rm tot}^{\eta-1}$. Here we take $\eta=1.0$ as the fiducial case, for which the weight is unity and the geometric kernel is recovered, and $\eta=1.5$ as the Euclidean source-count slope for a non-evolving population \citep{2010A&ARv..18....1D}, with an intermediate $\eta=1.2$. For the finite-core effect, which reduces the effective lensing probability by smoothing the high-magnification region when the source size is comparable to the lensing scale, the effective resolution is set by $\delta_{\rm eff}=\max(\delta,2\theta_{\rm core})$. Guided by the median VLBI core sizes of compact radio sources ($\sim$0.3 mas at 8 GHz and $\sim$1 mas at 2 GHz; \citealt{2012A&A...544A..34P}), we adopt $\theta_{\rm core}=0.3$~mas as a typical compact core at the search frequency and $\theta_{\rm core}=0.8$~mas as a conservative upper value. 
The ratios of the predicted $f_{\rm PBH}$ limits to the fiducial result $(\eta=1.0,\ \theta_{\rm core}=0)$ are shown in Fig.~\ref{fig:corrections}. Over $10^6$–$10^{9}\,M_\odot$, magnification bias tightens the limit by about 5\% for $\eta=1.2$ and about 13\% for $\eta=1.5$, relative to the fiducial case (rising to about 6\% and 16\% at $10^9\,M_\odot$). The finite-core effect acts in the opposite direction: when $2\theta_{\rm core}>\delta$, the source-core size sets $\delta_{\rm eff}$ and reduces the detectable cross section, most strongly at the low-mass end. For $\theta_{\rm core}=0.3~{\rm mas}, \delta_{\rm eff}=\delta$ and the result is almost unchanged. For $\theta_{\rm core}=0.8~{\rm mas}$, the limit is weakened by less than 1\% over the mass range where SMILE is most sensitive, becoming noticeable only near the low-mass edge, at about 20\% around $10^5\,M_\odot$, where the original constraint is already weak. For high-frequency compact-source searches, we take $\eta=1.2$ and $\theta_{\rm core}=0.3~{\rm mas}$ as values representative of the present sample. The core size remains below the resolution scale, so the combined result coincides with the $\eta=1.2$ case. In the current high-frequency search, both effects are generally negligible. However, the same framework is not limited to this regime. For lower-frequency imaging, e.g. around $2~{\rm GHz}$, $\theta_{\rm core}\sim1~{\rm mas}$ and $2\theta_{\rm core}$ can reach or exceed the resolution scale, so the low-mass cutoff would be mainly set by the finite-core size. A similar effect may also occur along strongly scattered, low-Galactic-latitude sightlines. Conversely, steeper source counts or higher-dynamic-range imaging with larger $R_{f,\max}$ would enhance the magnification bias. Thus, although these corrections are subdominant for the present prediction, they may become non-negligible for future surveys with different frequencies or higher sensitivities.

\begin{figure}
 \centering\includegraphics[width=\linewidth]{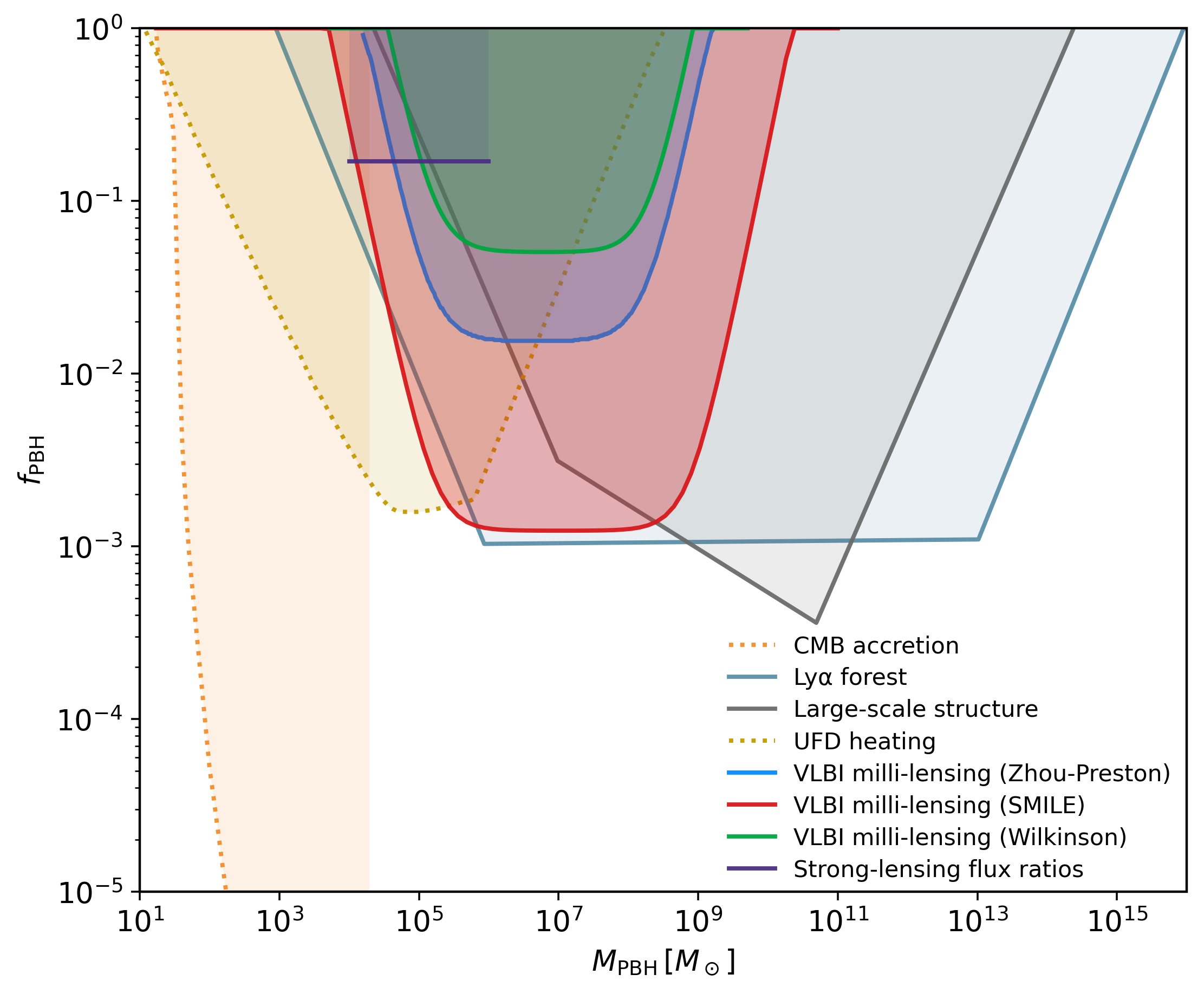}
 \caption{\textbf{Comparison of upper limits on the abundance of PBHs or compact-object dark matter candidates as a function of compact-object mass;} shaded regions above the curves are excluded, except the SMILE curve, which is a projected null-detection limit from this work. The fiducial SMILE VLBI milli-lensing prediction is compared with previous VLBI milli-lensing limits from \cite{2001PhRvL..86..584W,2022MNRAS.513.3627Z}. Other constraints shown for comparison include strong-lensing flux ratios \citep{2023MNRAS.522.5434D}, CMB accretion \citep{2020PhRvR...2b3204S}, Ly\(\alpha\) forest \citep{2026PhRvL.136q1402I}, large-scale structure \citep{2018MNRAS.478.3756C}, and UFD heating \citep{2024PhRvD.110g5011G}. Solid curves denote limits that are plotted here directly as upper limits on the PBH abundance; dotted curves denote complementary astrophysical or cosmological constraints that involve additional modelling assumptions or the interpretation of compact objects as PBHs. These constraints differ in physical origin and modeling assumptions, and are therefore shown as a broad, qualitative comparison across mass ranges.}
 \label{fig:compiled} 
\end{figure}

\textbf{Finally, we place our fiducial SMILE prediction in the broader context of several bounds on PBH abundance (Fig.~\ref{fig:compiled}).} Across the mass range compiled in Fig.~\ref{fig:compiled}, the available bounds originate from distinct physical mechanisms and are most sensitive in different mass regimes, so that no single probe dominates throughout. Among VLBI milli-lensing searches, SMILE yields lower $f_{\rm PBH}$ projected null-detection upper limits than \citet{2001PhRvL..86..584W} and \citet{2022MNRAS.513.3627Z} across the overlapping mass range, and extends to higher masses owing to its larger search window. In the massive regime probed here, VLBI milli-lensing provides one of the few geometric bounds, set by the lensing optical depth rather than by astrophysical modeling of structure formation or PBH accretion. As VLBI samples continue to grow in size and sensitivity, such milli-lensing predictions are expected to become an increasingly competitive component of the combined constraints on the PBH abundance.

\begin{figure}
 \centering
 \includegraphics[width=\linewidth]{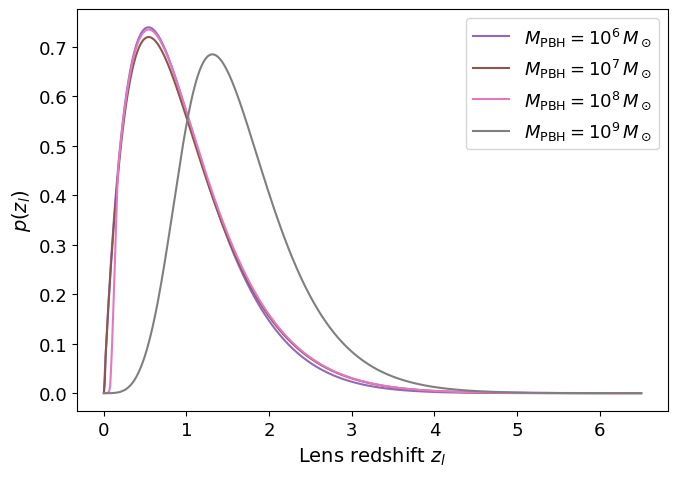}
 \caption{Normalized lens redshift distribution $p(z_l)$ as a function of lens redshift $z_l$, evaluated for four representative PBH masses $M_{\rm PBH} = 10^6,\ 10^7,\ 10^8,\ 10^9\,M_\odot$ using the SMILE observational parameters ($\delta=1.5\,\mathrm{mas}$, $\Delta=100\,\mathrm{mas}$, $R_{f,\rm max}=40$). The distribution $p(z_l)$ describes the relative contribution of lenses at different redshifts to the sample-averaged lensing optical depth.}
 \label{fig:pz_z}
\end{figure}

As shown above, the VLBI milli-lensing constraints are fundamentally shaped by the observational window, which selects a finite range of detectable image separations. This selection effect not only defines the mass range where the constraints are strongest, but also determines which part of the cosmic volume contributes most effectively to the optical depth. To quantify this, we differentiate the optical depth $\bar{\tau}$ (the sample-averaged optical depth defined above) with respect to lens redshift and define a normalized lens redshift distribution
\[
p(z_L)=\frac{1}{\bar{\tau}}\frac{d\bar{\tau}}{dz_L},
\]
which describes the relative contribution of lenses at different redshifts. We further define an effective lens redshift
\[
z_{\rm eff}=\int z_L\,p(z_L)\,dz_L,
\]
as a compact characterization of the typical redshift scale probed by VLBI milli-lensing.
We find that for $M_{\rm PBH}\sim10^6$--$10^8\,M_\odot$, the distribution $p(z_L)$ peaks at $z_L\sim0.4$--$0.5$ and rapidly declines at both lower and higher redshift (see Fig.~\ref{fig:pz_z}), indicating that the dominant contribution arises from a relatively narrow redshift interval. At the high-mass end, the peak shifts to higher redshift (e.g. the $10^9 M_\odot$ curve), since $M_{\rm PBH}$ enters only through the selection kernel in Eq.~(\ref{eq:ybounds}). A larger $\theta_E$ pushes the image separation at low $z_L$ above the search limit $\Delta$, removing the low-redshift contribution and shifting the peak to $z_L\sim1.3$. Correspondingly, the effective redshift $z_{\rm eff}$ remains nearly constant at $\sim1$ in the mass range $10^6$--$10^8\,M_\odot$, and increases moderately at higher masses. This behavior is governed by the structure of the lensing kernel $\frac{D_L D_{LS}}{D_S}\cdot\frac{(1+z_L)^2}{H(z_L)}$: at low redshift the geometric factor is not yet fully developed, while at high redshift the growth of $H(z)$ suppresses the contribution. 
In addition, the VLBI selection condition \(\delta_{\rm eff} < \Delta\theta < \Delta\) further restricts the allowed lens configurations, effectively selecting a limited redshift range that shifts slowly with PBH mass. The central $80\%$ of the contribution arises from $z_L \simeq 0.3\text{--}2.0$ for $10^6\text{--}10^8\,M_\odot$, and from $z_L \simeq 0.9\text{--}2.5$ for $10^9\,M_\odot$. After marginalizing over the source-redshift distribution, $p(z_L)$ is mainly set by the lensing geometry and the VLBI selection function, and is only weakly sensitive to the PBH mass function. Thus, even for the highest-redshift sources in the sample, the dominant lens contribution remains confined to this interval. The predicted limits therefore reflect the PBH abundance within this window rather than along the full line of sight to $z_S$.

Given that the effective lens redshift is $z_{\rm eff}\sim1$, with most of the weight lying below $z_L \sim 2$, an important question is whether PBHs can experience significant mass growth in this redshift range, and thus modify the inferred constraints. In the analysis above, we have assumed that PBH masses remain constant. However, PBHs may grow through accretion of surrounding baryonic matter. While rapid accretion is generally expected at very high redshift ($z\gtrsim20$), the efficiency of this process decreases substantially at later times due to cosmic expansion and the declining ambient gas density. Nevertheless, PBHs may trace the underlying dark matter distribution and become embedded in virialized halos during structure formation, where the ambient gas density can be significantly enhanced \cite[][]{2024ApJ...975..139Z}. In such relatively dense environments, residual accretion could in principle affect the lens mass and hence the lensing signal. To estimate the maximal impact of this effect, we adopt a commonly used Bondi–Hoyle–Lyttleton (BHL) accretion model \citep[]{2004NewAR..48..843E}. The accretion rate is given by
\[
\dot{M} = 4\pi\lambda \rho G^2 M^2 / v_{\rm eff}^3,
\]
where the characteristic scale is set by the Bondi radius $r_B = GM/v_{\rm eff}^2$, and the effective velocity $v_{\rm eff}=\sqrt{v_{\rm rel}^2+c_s^2}$ combines the PBH–gas relative velocity $v_{\rm rel}$ and the gas sound speed $c_s$. The dimensionless parameter $\lambda$ encodes the dependence of the accretion rate on the flow geometry and thermodynamic state. A key uncertainty in modeling PBH accretion is the surrounding baryonic environment. In standard treatments—particularly those related to CMB constraints—the gas is often assumed to follow the cosmic mean density, which is appropriate for describing global energy injection \citep{2008ApJ...680..829R,2017PhRvD..95d3534A}. However, for gravitational lensing by individual compact objects, the local environment may differ substantially from the cosmic average. If PBHs trace the dark matter distribution, they may reside in virialized halos formed during structure formation \citep{2016PhRvD..94h3504C,2018CQGra..35f3001S}. In such systems, especially those with virial temperatures $T_{\rm vir}\gtrsim10^4\,\mathrm{K}$, baryonic gas can efficiently cool and remain bound, providing a reservoir for accretion \citep{2008ApJ...680..829R}. 

In this work, we do not attempt to model the full distribution of PBH environments. Instead, we adopt an idealized but physically motivated scenario in which PBHs reside in gas-rich virialized halos, in order to estimate the maximal impact of accretion on the lensing constraints. Under this assumption, the baryon density is parameterized as $\rho_b(z)=\rho_0(1+z)^3$, with $\rho_0=\Delta_b\,\bar{\rho}_{b,0}$ and $\Delta_b\simeq200$ representing a typical virial overdensity, corresponding to $\rho_0\simeq1.2\times10^{12}\,M_\odot\,\mathrm{Mpc}^{-3}$ \citep{1998ApJ...495...80B}. For the thermodynamic state of the gas, we adopt a characteristic virial temperature $T_{\rm vir}\sim10^4\,\mathrm{K}$, yielding a sound speed $c_s\simeq15\,\mathrm{km\,s^{-1}}$ \citep{2008ApJ...680..829R}. The relative velocity is taken to be of order the halo virial velocity, $v_{\rm rel}\sim17\,\mathrm{km\,s^{-1}}$, leading to an effective velocity $v_{\rm eff}\simeq22\,\mathrm{km\,s^{-1}}$. For the accretion parameter, we adopt a fiducial value $\lambda=0.6$, which lies between the adiabatic limit ($\lambda\approx0.25$ for $\gamma=5/3$) and the isothermal limit ($\lambda\approx1.12$ for $\gamma=1$) of the classical Bondi solution \citep{1952MNRAS.112..195B}, and represents a reasonable intermediate case for gas in a virialized halo environment. This setup serves as a controlled and optimistic scenario to assess whether accretion-driven mass growth could significantly modify the VLBI milli-lensing constraints.

\begin{figure}
 \centering
 \includegraphics[width=\linewidth]{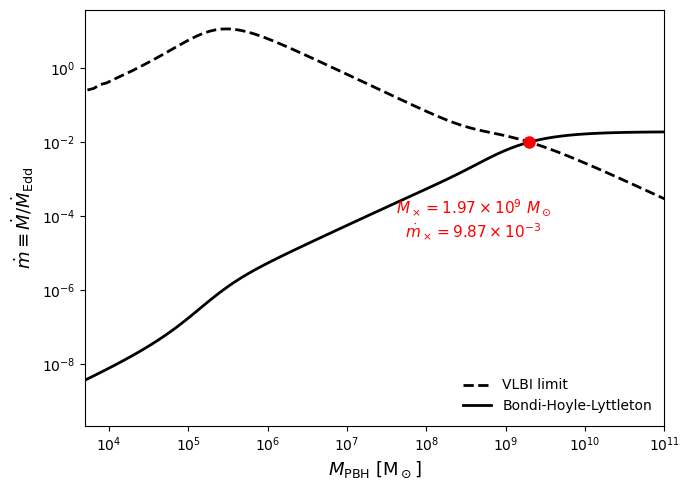}
 \caption{Eddington ratio $\dot{m} \equiv \dot{M}/\dot{M}_{\rm Edd}$ as a function of PBH mass, evaluated at the effective lens redshift $z_{\rm eff}(M)$ characteristic of the lensing optical depth. The dashed line shows the maximum $\dot{m}$ allowed by the VLBI flux upper limit; the solid line shows the Bondi--Hoyle--Lyttleton rate for the fiducial virialized halo parameters described in the text. The red dot marks the critical mass $M_\times$ above which BHL accretion would violate the VLBI constraint.}
 \label{fig:L_acc}
\end{figure}

Another requirement is that any radio emission produced by PBH accretion in the lens plane must remain undetected in the VLBA images used for the milli-lensing search. Such emission would appear as an additional compact component superposed on the background radio source image. Since the SMILE search is based on VLBA imaging, we require the accretion flux density to be below a $3\sigma_{\rm rms}$ detection threshold for an additional compact component. Typical VLBA image rms values are of order $0.1\text{--}1 \, \mathrm{mJy}$ and we adopt the conservative representative value $\sigma_{\rm rms}=1 \, \mathrm{mJy/beam}$, giving $F_{\nu,\rm acc}^{\rm max}=3\,\mathrm{mJy}$. For an accreting PBH at the effective lens redshift $z_{\rm eff}$, this corresponds to the characteristic luminosity limit
$
L_{\rm acc}^{\rm max} \simeq 4\pi D_{\rm lum}^2(z_{\rm eff})\nu F_{\nu,\rm acc}^{\rm max},
$
where $D_{\rm lum}(z_{\rm eff})$ is the luminosity distance to the effective lens redshift, and $\nu=5\,\mathrm{GHz}$ is taken as the representative VLBA C-band imaging frequency. The factor $\nu F_\nu$ provides an order-of-magnitude, upper estimate of the accretion luminosity. The maximum allowed Eddington ratio is then $\dot{m}_{\rm max} = \frac{L_{\rm acc}^{\rm max}}{\epsilon\, L_{\rm Edd}}$, where $L_{\rm Edd} = 1.26\times10^{38}(M/M_\odot)\,\mathrm{erg\,s^{-1}}$ and we adopt a radiative efficiency $\epsilon = 0.1$. Both the accretion rate and the luminosity limit are evaluated at the mass-dependent effective lens redshift $z_{\rm eff}(M)$, so that each PBH mass is compared at the redshift that dominates its lensing contribution. As shown in Fig.~\ref{fig:L_acc}, we find that for $M\sim10^6\text{--}10^9\,M_\odot$, the BHL accretion rate remains below this observational limit, implying that even in such dense environments the accretion luminosity stays below the detection threshold and would not appear as a separate component. Only at higher masses does the luminosity constraint begin to impose a meaningful restriction. Specifically, the BHL accretion rate exceeds the VLBI flux upper limit above a critical mass $M_\times \simeq 1.97\times10^{9}\,M_\odot$, at which the Eddington ratio reaches $\dot{m}_\times \simeq 9.87\times10^{-3}$. 
Below this threshold, the BHL rate lies comfortably beneath the observational limit, meaning that accretion in this regime would be radiatively sub-threshold and undetectable by current VLBI surveys. Above \(M_\times\), the assumed BHL accretion rate would produce radio emission in excess of the VLBI survey detection limit, rendering the strong-accretion scenario observationally disfavored.

Since $M_\times$ lies above the mass range where the milli-lensing constraints are most sensitive, accretion is radiatively undetectable throughout the VLBI-sensitive window, and the inferred $f_{\rm PBH}$ limits are essentially unchanged: including accretion changes the limits within this window at the sub-percent level. At higher masses, accretion could in principle drive appreciable mass growth. This regime, however, lies outside the VLBI-sensitive window, and the strong-accretion scenario is in any case disfavored by the VLBI non-detection above $M_\times$. Therefore, under the adopted halo-accretion scenario, PBH mass growth has little impact on the predicted milli-lensing constraints, which are set by the lensing geometry and the VLBI selection function rather than by the accretion history of the PBHs.

\section{Conclusion}
\label{sec:Conclusion}

In this paper, we have presented predictions for future constraints on the abundance of primordial black holes, $f_{\rm PBH}$, in the mass range $M_{\rm PBH}\sim10^6\text{--}10^9\,M_\odot$ using the VLBI milli-lensing method. By considering the SMILE sample of 4968 flat-spectrum compact radio sources together with the redshift distribution given in \citet{2022A&A...668A.166L}, we obtain predicted limits with a characteristic shape determined by the VLBI observational window. The strongest limits arise at intermediate masses where the Einstein radius falls within the detectable separation range, and weaken toward both low and high masses as the image separation moves outside the observational window. Compared with previous studies, the SMILE sample yields stronger predicted limits due to its larger number of sources and wider effective search region. For a zero-detection scenario with $\delta=1.5\,\mathrm{mas}$, $\Delta=100\,\mathrm{mas}$, and $R_{f,\max}=40$, the strongest predicted 95\% upper limit reaches $f_{\rm PBH}\lesssim0.12\%$ near $M_{\rm PBH}\sim\mathcal{O}(10^7)\,M_\odot$, approximately an order of magnitude below previous VLBI milli-lensing limits. Consistent with standard lensing statistics, the nearly order-of-magnitude increase in source number has a greater impact on the predicted limits than increasing $R_{f,\max}$ alone. Allowing extended mass functions introduces an additional model dependence: broader lognormal distributions shift more probability outside the VLBI-sensitive mass window and therefore yield weaker predicted limits than the monochromatic case. The statistical interpretation is likewise sensitive to the number of confirmed events. Owing to the rare-event nature of milli-lensing, a single confirmed event among the 4968 SMILE sources would raise the predicted upper limit from $0.12\%$ to $0.20\%$.

The predicted limits are robust against several sources of systematic uncertainty. Magnification bias increases the effective lensing probability, whereas a finite source core reduces the resolvable cross section when $2\theta_{\rm core}>\delta$. For the representative SMILE choices $\eta=1.2$ and $\theta_{\rm core}=0.3\,\mathrm{mas}$, the finite-core correction is inactive, while magnification bias tightens the predicted limits by about $5\%$. Beyond these source-level corrections, the VLBI selection window also determines which lens redshifts dominate the optical depth. For $M_{\rm PBH}\sim10^6\text{--}10^8\,M_\odot$, the balance between lensing geometry and the image-separation cuts places the peak at $z_L\sim0.4\text{--}0.5$. At $10^9\,M_\odot$, the larger Einstein radius moves low-redshift image pairs beyond the maximum search separation, shifting the peak to $z_L\sim1.3$. The predicted limits therefore primarily reflect the PBH abundance within this selection-defined lens-redshift interval. Because most of the lensing weight lies below $z_L\sim2$, we also assessed whether PBH accretion over this interval could significantly change the lens mass and hence the predicted limits. Even under the optimistic halo-accretion scenario, including PBH mass growth changes the predicted limits by less than $1\%$ within the VLBI-sensitive window. More significant mass growth is possible only at higher masses, beyond the most sensitive window. However, the strong-accretion scenario would exceed the VLBI detection threshold above $1.97\times10^{9}\,M_\odot$ and is therefore observationally disfavored. The predictions presented here are also complementary to scenarios motivated by JWST observations of massive galaxies at $z\gtrsim7$: while PBHs that serve as black hole seeds in dense environments are expected to grow well beyond this mass window by the present day, our results provide predicted limits on the abundance of the remaining population of isolated PBHs that have not undergone significant accretion.

In the future, the most direct path to improving VLBI milli-lensing constraints on PBHs lies in two aspects. From an observational perspective, the most significant progress will be driven by larger source catalogs. In particular, the upcoming wide-field radio surveys such as those planned with the next-generation Very Large Array (ngVLA) and the Square Kilometre Array (SKA-Mid) will dramatically increase the number of detectable flat-spectrum compact radio sources, potentially extending statistical samples to $\sim10^5$ sources \citep{2015aska.confE.143P,Selina:2024}. Equally important is the confirmation of candidate milli-lensing systems through dedicated multi-frequency and multi-epoch follow-up observations. Even a single robustly confirmed event would qualitatively change the nature of the inference by moving it beyond the null-detection regime. The strong sensitivity to the number of confirmed events highlights the importance of a more rigorous statistical treatment in future analyses. From a statistical perspective, a natural extension would be to develop a hierarchical Bayesian framework that can formally incorporate the partial confirmation status of milli-lensing candidates. A more principled treatment would assign each candidate system a classification probability $p_i$ derived from multi-frequency follow-up criteria, such as flux-ratio stability, spectral index consistency, and morphology, and incorporate these probabilities directly into the likelihood function. This would replace such discrete scenarios with a formal posterior on $f_{\mathrm{PBH}}$, naturally interpolating between the zero-detection and confirmed-detection limits. Such a framework would be particularly valuable as future VLBI surveys accumulate larger candidate samples, where the number of partially confirmed systems may become the dominant source of statistical uncertainty. Finally, combining VLBI milli-lensing with complementary probes -- CMB constraints at higher masses and gravitational wave observations at stellar masses -- will enable a more complete reconstruction of the PBH mass function across different mass scales.%

\section*{Acknowledgements}

This work is supported by Beijing Natural Science Foundation No. 1242021 and the National Natural Science Foundation of China under Grants No. 12433001.

\bibliography{ref}{}

@ARTICLE{2020A&A...641A...6P,
       author = {{Aghanim}, N. and {Akrami}, Y. and et al.},
        title = "{Planck 2018 results. VI. Cosmological parameters}",
      journal = {\text{A\&A}},
         year = 2020,
        month = sep,
       volume = {641},
          eid = {A6},
        pages = {A6},
       adsurl = {https://ui.adsabs.harvard.edu/abs/2020A&A...641A...6P}
}

@ARTICLE{1980ApJ...238..471R,
       author = {{Rubin}, V.~C. and {Ford}, W.~K., Jr. and {Thonnard}, N.},
        title = "{Rotational properties of 21 Sc galaxies with a large range of luminosities and radii, from NGC 4605 /R = 4kpc/ to UGC 2885 /R = 122kpc/.}",
      journal = {\text{ApJ}},
         year = 1980,
        month = jun,
       volume = {238},
        pages = {471-487},
       adsurl = {https://ui.adsabs.harvard.edu/abs/1980ApJ...238..471R}
}

@ARTICLE{1974MNRAS.168..399C,
       author = {{Carr}, B.~J. and {Hawking}, S.~W.},
        title = "{Black holes in the early Universe}",
      journal = {\text{MNRAS}},
         year = 1974,
        month = aug,
       volume = {168},
        pages = {399-416},
       adsurl = {https://ui.adsabs.harvard.edu/abs/1974MNRAS.168..399C}
}

@ARTICLE{2016PhRvD..94h3504C,
       author = {{Carr}, Bernard and {K{\"u}hnel}, Florian and {Sandstad}, Marit},
        title = "{Primordial black holes as dark matter}",
      journal = {\text{PRD}},
         year = 2016,
        month = oct,
       volume = {94},
       number = {8},
          eid = {083504},
        pages = {083504},
       adsurl = {https://ui.adsabs.harvard.edu/abs/2016PhRvD..94h3504C}
}

@ARTICLE{2010PhRvD..81j4019C,
       author = {{Carr}, B.~J. and {Kohri}, K. and {Sendouda}, Y. and {Yokoyama}, J.},
        title = "{New cosmological constraints on primordial black holes}",
      journal = {\text{PRD}},
         year = 2010,
        month = may,
       volume = {81},
       number = {10},
          eid = {104019},
        pages = {104019},
       adsurl = {https://ui.adsabs.harvard.edu/abs/2010PhRvD..81j4019C}
}

@ARTICLE{2021JPhG...48d3001G,
       author = {{Green}, Anne M. and {Kavanagh}, Bradley J.},
        title = "{Primordial black holes as dark matter: recent developments}",
      journal = {Journal of Physics G Nuclear Physics},
         year = 2021,
        month = apr,
       volume = {48},
       number = {4},
          eid = {043001},
        pages = {043001},
       adsurl = {https://ui.adsabs.harvard.edu/abs/2021JPhG...48d3001G}
}

@ARTICLE{1952MNRAS.112..195B,
       author = {{Bondi}, H.},
        title = "{On spherically symmetric accretion}",
      journal = {\text{MNRAS}},
         year = 1952,
        month = jan,
       volume = {112},
        pages = {195},
       adsurl = {https://ui.adsabs.harvard.edu/abs/1952MNRAS.112..195B}
}

@ARTICLE{2017PhRvD..95d3534A,
       author = {{Ali-Ha{\"\i}moud}, Yacine and {Kamionkowski}, Marc},
        title = "{Cosmic microwave background limits on primordial black holes with early primordial spectrum asymmetries}",
      journal = {\text{PRD}},
         year = 2017,
        month = feb,
       volume = {95},
       number = {4},
          eid = {043534},
        pages = {043534},
       adsurl = {https://ui.adsabs.harvard.edu/abs/2017PhRvD..95d3534A}
}

@ARTICLE{2024SCPMA..6789511T,
       author = {{Tu}, Zi-Liang and {Chen}, Yu-Xuan and {Liu}, Xiao-Jun and {Wang}, Bo and {Gao}, Yu and {Zhang}, Xin},
        title = "{Rapidly growing primordial black holes as seeds of the massive high-redshift JWST Galaxies}",
      journal = {\text{SCPMA}},
         year = 2024,
        month = aug,
       volume = {67},
       number = {10},
          eid = {109512},
        pages = {109512},
}

@ARTICLE{2022ApJ...940L..14N,
       author = {{Naidu}, Rohan P. and {Oesch}, Pascal A. and {van Dokkum}, Pieter and {Nelson}, Erica J. and {Suess}, Katherine A. and {Brammer}, Gabriel and {Whitaker}, Katherine E. and {Illingworth}, Garth and {Bouwens}, Rychard J. and {Tacchella}, Sandro and et al.},
        title = "{Two Remarkably Luminous Galaxy Candidates at z {\ensuremath{\approx}} 10-12 Revealed by JWST}",
      journal = {\text{ApJL}},
         year = 2022,
        month = nov,
       volume = {940},
       number = {1},
          eid = {L14},
        pages = {L14},
       adsurl = {https://ui.adsabs.harvard.edu/abs/2022ApJ...940L..14N}
}

@ARTICLE{2022ApJ...938L..15C,
       author = {{Castellano}, M. and {Fontana}, A. and {Treu}, T. and {Santini}, P. and {Merlin}, E. and {Leethochawalit}, N. and {Trenti}, M. and {Pentericci}, L. and {Zavala}, J. and {Calabr{\`o}}, A. and et al.},
        title = "{Early Results from JWST: The Galaxy Luminosity Function at z {\ensuremath{\approx}} 10-17}",
      journal = {\text{ApJL}},
         year = 2022,
        month = oct,
       volume = {938},
       number = {2},
          eid = {L15},
        pages = {L15},
       adsurl = {https://ui.adsabs.harvard.edu/abs/2022ApJ...938L..15C}
}

@ARTICLE{2025JCAP...04..040Z,
       author = {{Ziparo}, F. and {Gallerani}, S. and {Ferrara}, A.},
        title = "{Primordial black holes as supermassive black hole seeds}",
      journal = {\jcap},
         year = 2025,
        month = apr,
       volume = {2025},
       number = {4},
          eid = {040},
        pages = {040},
 primaryClass = {astro-ph.CO},
       adsurl = {https://ui.adsabs.harvard.edu/abs/2025JCAP...04..040Z}
}

@ARTICLE{2008ApJ...680..829R,
       author = {{Ricotti}, Massimo and {Ostriker}, Jeremiah P. and {Mack}, Katherine J.},
        title = "{Effect of Primordial Black Holes on the Cosmic Microwave Background and Cosmological Parameter Extraction}",
      journal = {\text{ApJ}},
         year = 2008,
        month = jun,
       volume = {680},
       number = {2},
        pages = {829-845},
       adsurl = {https://ui.adsabs.harvard.edu/abs/2008ApJ...680..829R}
}

@ARTICLE{2001PhRvL..86..584W,
       author = {{Wilkinson}, P.~N. and {Henstock}, D.~R. and {Browne}, I.~W.~A. and {Polatidis}, A.~G. and {Augusto}, P. and {Readhead}, A.~C.~S. and {Pearson}, T.~J. and {Xu}, W. and {Taylor}, G.~B.},
        title = "{Limits on the Cosmological Abundance of Compact Objects in the Mass Range 10^{6}-10^{8} M_{\odot}}",
      journal = {\text{PRL}},
         year = 2001,
        month = jan,
       volume = {86},
       number = {4},
        pages = {584-587},
       adsurl = {https://ui.adsabs.harvard.edu/abs/2001PhRvL..86..584W}
}

@ARTICLE{2022MNRAS.513.3627Z,
       author = {{Zhou}, Huan and {Lian}, Yujie and {Li}, Zhengxiang and {Cao}, Shuo and {Huang}, Zhiqi},
        title = "{Constraints on the abundance of supermassive primordial black holes from lensing of compact radio sources}",
      journal = {\mnras},
         year = 2022,
        month = jul,
       volume = {513},
       number = {3},
        pages = {3627-3633},
       adsurl = {https://ui.adsabs.harvard.edu/abs/2022MNRAS.513.3627Z}
}

@ARTICLE{2017PhRvD..96h3524P,
       author = {{Poulin}, Vivian and {Serpico}, Pasquale D. and {Calore}, Francesca and {Clesse}, S{\'e}bastien and {Kohri}, Kazunori},
        title = "{CMB bounds on disk-accreting massive primordial black holes}",
      journal = {\prd},
         year = 2017,
        month = oct,
       volume = {96},
       number = {8},
          eid = {083524},
        pages = {083524},
       adsurl = {https://ui.adsabs.harvard.edu/abs/2017PhRvD..96h3524P}
}

@ARTICLE{1985AJ.....90.1599P,
       author = {{Preston}, R.~A. and {Morabito}, D.~D. and {Williams}, J.~G. and {Faulkner}, J. and {Jauncey}, D.~L. and {Nicolson}, G.},
        title = "{A VLBI survey at 2.29 GHz.}",
      journal = {\aj},
         year = 1985,
        month = sep,
       volume = {90},
        pages = {1599-1603},
       adsurl = {https://ui.adsabs.harvard.edu/abs/1985AJ.....90.1599P}
}

@ARTICLE{2006JCAP...11..002J,
       author = {{Jackson}, J.~C. and {Jannetta}, A.~L.},
        title = "{Legacy data and cosmological constraints from the angular-size/redshift relation for ultracompact radio sources}",
      journal = {\jcap},
         year = 2006,
        month = nov,
       volume = {2006},
       number = {11},
          eid = {002},
        pages = {002},
       adsurl = {https://ui.adsabs.harvard.edu/abs/2006JCAP...11..002J}
}

@ARTICLE{2021MNRAS.507L...6C,
       author = {{Casadio}, C. and {Blinov}, D. and {Readhead}, A.~C.~S. and {Browne}, I.~W.~A. and {Wilkinson}, P.~N. and {Hovatta}, T. and {Mandarakas}, N. and {Pavlidou}, V. and {Tassis}, K. and {Vedantham}, H.~K. and {Zensus}, J.~A. and {Diamantopoulos}, V. and {Dolapsaki}, K.~E. and {Gkimisi}, K. and {Kalaitzidakis}, G. and {Mastorakis}, M. and {Nikolaou}, K. and {Ntormousi}, E. and {Pelgrims}, V. and {Psarras}, K.},
        title = "{SMILE: Search for MIlli-LEnses}",
      journal = {\mnras},
         year = 2021,
        month = oct,
       volume = {507},
       number = {1},
        pages = {L6-L10},
       adsurl = {https://ui.adsabs.harvard.edu/abs/2021MNRAS.507L...6C}
}

@ARTICLE{2022A&A...668A.166L,
       author = {{Loudas}, Nick and {Pavlidou}, Vasiliki and {Casadio}, Carolina and {Tassis}, Konstantinos},
        title = "{Discriminating power of milli-lensing observations for dark matter models}",
      journal = {\aap},
         year = 2022,
        month = dec,
       volume = {668},
          eid = {A166},
        pages = {A166},
       adsurl = {https://ui.adsabs.harvard.edu/abs/2022A&A...668A.166L}
}

@ARTICLE{2003MNRAS.341....1M,
       author = {{Myers}, S.~T. and {Jackson}, N.~J. and {Browne}, I.~W.~A. and {de Bruyn}, A.~G. and {Pearson}, T.~J. and {Readhead}, A.~C.~S. and {Wilkinson}, P.~N. and {Biggs}, A.~D. and {Blandford}, R.~D. and {Fassnacht}, C.~D. and {Koopmans}, L.~V.~E. and {Marlow}, D.~R. and {McKean}, J.~P. and {Norbury}, M.~A. and {Phillips}, P.~M. and {Rusin}, D. and {Shepherd}, M.~C. and {Sykes}, C.~M.},
        title = "{The Cosmic Lens All-Sky Survey - I. Source selection and observations}",
      journal = {\mnras},
         year = 2003,
        month = may,
       volume = {341},
       number = {1},
        pages = {1-12},
       adsurl = {https://ui.adsabs.harvard.edu/abs/2003MNRAS.341....1M}
}

@ARTICLE{2025ApJS..276...38P,
       author = {{Petrov}, L.~Y. and {Kovalev}, Y.~Y.},
        title = "{The Radio Fundamental Catalog. I. Astrometry}",
      journal = {\apjs},
         year = 2025,
        month = feb,
       volume = {276},
       number = {2},
          eid = {38},
        pages = {38},
       adsurl = {https://ui.adsabs.harvard.edu/abs/2025ApJS..276...38P}
}

@ARTICLE{2025A&A...695A.169P,
       author = {{P{\"o}tzl}, F.~M. and {Casadio}, C. and {Kalaitzidakis}, G. and {{\'A}lvarez-Ortega}, D. and {Kumar}, A. and {Missaglia}, V. and {Blinov}, D. and {Janssen}, M. and {Loudas}, N. and {Pavlidou}, V. and {Readhead}, A.~C.~S. and {Tassis}, K. and {Wilkinson}, P.~N. and {Zensus}, J.~A.},
        title = "{SMILE: Discriminating milli-lens systems in a VLBI pilot project}",
      journal = {\aap},
         year = 2025,
        month = mar,
       volume = {695},
          eid = {A169},
        pages = {A169},
       adsurl = {https://ui.adsabs.harvard.edu/abs/2025A&A...695A.169P}
}

@ARTICLE{1984ApJ...284....1T,
       author = {{Turner}, E.~L. and {Ostriker}, J.~P. and {Gott}, III, J.~R.},
        title = "{The statistics of gravitational lenses : the distributions of image angular separations and lens redshifts.}",
      journal = {\apj},
         year = 1984,
        month = sep,
       volume = {284},
        pages = {1-22},
       adsurl = {https://ui.adsabs.harvard.edu/abs/1984ApJ...284....1T}
}

@ARTICLE{2017JCAP...09..037R,
       author = {{Raidal}, Martti and {Vaskonen}, Ville and {Veerm{\"a}e}, Hardi},
        title = "{Gravitational waves from primordial black hole mergers}",
      journal = {\jcap},
         year = 2017,
        month = sep,
       volume = {2017},
       number = {9},
          eid = {037},
        pages = {037},
       adsurl = {https://ui.adsabs.harvard.edu/abs/2017JCAP...09..037R}
}

@ARTICLE{2017PhRvD..96b3514C,
       author = {{Carr}, Bernard and {Raidal}, Martti and {Tenkanen}, Tommi and {Vaskonen}, Ville and {Veerm{\"a}e}, Hardi},
        title = "{Primordial black hole constraints for extended mass functions}",
      journal = {PRD},
         year = 2017,
        month = jul,
       volume = {96},
       number = {2},
          eid = {023514},
        pages = {023514},
       adsurl = {https://ui.adsabs.harvard.edu/abs/2017PhRvD..96b3514C}
}

@ARTICLE{2019EPJC...79..717L,
       author = {{Liu}, Lang and {Guo}, Zong-Kuan and {Cai}, Rong-Gen},
        title = "{Effects of the merger history on the merger rate density of primordial black hole binaries}",
      journal = {EPJC},
         year = 2019,
        month = aug,
       volume = {79},
       number = {8},
          eid = {717},
        pages = {717},
       adsurl = {https://ui.adsabs.harvard.edu/abs/2019EPJC...79..717L}
}

@ARTICLE{2012ApJ...755...31C,
       author = {{Cao}, Shuo and {Covone}, Giovanni and {Zhu}, Zong-Hong},
        title = "{Testing the Dark Energy with Gravitational Lensing Statistics}",
      journal = {\apj},
         year = 2012,
        month = aug,
       volume = {755},
       number = {1},
          eid = {31},
        pages = {31}
}

@ARTICLE{2012JCAP...03..016C,
       author = {{Cao}, Shuo and {Pan}, Yu and {Biesiada}, Marek and {Godlowski}, Wlodzimierz and {Zhu}, Zong-Hong},
        title = "{Constraints on cosmological models from strong gravitational lensing systems}",
      journal = {\jcap},
         year = 2012,
        month = mar,
       volume = {2012},
       number = {3},
          eid = {016},
        pages = {016}
}

@ARTICLE{2015ApJ...806..185C,
       author = {{Cao}, Shuo and {Biesiada}, Marek and {Gavazzi}, Rapha{\"e}l and {Pi{\'o}rkowska}, Aleksandra and {Zhu}, Zong-Hong},
        title = "{Cosmology with Strong-lensing Systems}",
      journal = {\apj},
         year = 2015,
        month = jun,
       volume = {806},
       number = {2},
          eid = {185},
        pages = {185}
}

@ARTICLE{2015ApJ...806...66C,
       author = {{Cao}, Shuo and {Biesiada}, Marek and {Zheng}, Xiaogang and {Zhu}, Zong-Hong},
        title = "{Exploring the Properties of Milliarcsecond Radio Sources}",
      journal = {\apj},
         year = 2015,
        month = jun,
       volume = {806},
       number = {1},
          eid = {66},
        pages = {66}
}

@ARTICLE{2018CQGra..35f3001S,
       author = {{Sasaki}, Misao and {Suyama}, Teruaki and {Tanaka}, Takahiro and {Yokoyama}, Shuichiro},
        title = "{Primordial black holes{\textemdash}perspectives in gravitational wave astronomy}",
      journal = {CQG},
         year = 2018,
        month = mar,
       volume = {35},
       number = {6},
          eid = {063001},
        pages = {063001},
       adsurl = {https://ui.adsabs.harvard.edu/abs/2018CQGra..35f3001S}
}

@ARTICLE{2004NewAR..48..843E,
       author = {{Edgar}, Richard},
        title = "{A review of Bondi-Hoyle-Lyttleton accretion}",
      journal = {\nar},
         year = 2004,
        month = sep,
       volume = {48},
       number = {10},
        pages = {843-859},
       adsurl = {https://ui.adsabs.harvard.edu/abs/2004NewAR..48..843E}
}

@ARTICLE{1998ApJ...495...80B,
       author = {{Bryan}, Greg L. and {Norman}, Michael L.},
        title = "{Statistical Properties of X-Ray Clusters: Analytic and Numerical Comparisons}",
      journal = {\apj},
         year = 1998,
        month = mar,
       volume = {495},
       number = {1},
        pages = {80-99},
       adsurl = {https://ui.adsabs.harvard.edu/abs/1998ApJ...495...80B}
}

@INPROCEEDINGS{Selina:2024,
       author = {{Selina}, Robert and {Murphy}, Eric and {Beasley}, Anthony},
        title = "{The ngVLA: A Technical Overview}",
    booktitle = {American Astronomical Society Meeting Abstracts \#241},
         year = 2023,
       volume = {241},
        month = jan,
          eid = {357.02},
        pages = {357.02},
       adsurl = {https://ui.adsabs.harvard.edu/abs/2023AAS...24135702S}
}

@ARTICLE{2017A&A...606A..15C,
       author = {{Cao}, Shuo and {Zheng}, Xiaogang and {Biesiada}, Marek and {Qi}, Jingzhao and {Chen}, Yun and {Zhu}, Zong-Hong},
        title = "{Ultra-compact structure in intermediate-luminosity radio quasars: building a sample of standard cosmological rulers and improving the dark energy constraints up to z   3}",
      journal = {A\&A},
         year = 2017,
        month = sep,
       volume = {606},
          eid = {A15},
        pages = {A15},
       adsurl = {https://ui.adsabs.harvard.edu/abs/2017A&A...606A..15C}
}

@ARTICLE{2017JCAP...02..012C,
       author = {{Cao}, Shuo and {Biesiada}, Marek and {Jackson}, John and {Zheng}, Xiaogang and {Zhao}, Yuhang and {Zhu}, Zong-Hong},
        title = "{Measuring the speed of light with ultra-compact radio quasars}",
      journal = {JCAP},
         year = 2017,
        month = feb,
       volume = {2017},
       number = {2},
          eid = {012},
        pages = {012},
       adsurl = {https://ui.adsabs.harvard.edu/abs/2017JCAP...02..012C}
}

@ARTICLE{1973ApJ...185..397P,
       author = {{Press}, William H. and {Gunn}, James E.},
        title = "{Method for Detecting a Cosmological Density of Condensed Objects}",
      journal = {\apj},
         year = 1973,
        month = oct,
       volume = {185},
        pages = {397-412}
}

@ARTICLE{2025JCAP...04..023G,
       author = {{Green}, Anne M.},
        title = "{Primordial black hole stellar microlensing constraints: understanding their dependence on the density and velocity distributions}",
      journal = {\jcap},
         year = 2025,
        month = apr,
       volume = {2025},
       number = {4},
          eid = {023},
        pages = {023}
}

@ARTICLE{2024CQGra..41n3001D,
       author = {{Dom{\`e}nech}, Guillem and {Sasaki}, Misao},
        title = "{Probing primordial black hole scenarios with terrestrial gravitational wave detectors}",
      journal = {Classical and Quantum Gravity},
         year = 2024,
        month = jul,
       volume = {41},
       number = {14},
          eid = {143001},
        pages = {143001}
}

@ARTICLE{2022MNRAS.511.1141Z,
       author = {{Zhou}, Huan and {Li}, Zhengxiang and {Huang}, Zhiqi and {Gao}, He and {Huang}, Lu},
        title = "{Constraints on the abundance of primordial black holes with different mass distributions from lensing of fast radio bursts}",
      journal = {\mnras},
         year = 2022,
        month = mar,
       volume = {511},
       number = {1},
        pages = {1141-1152}
}

@ARTICLE{2018PhRvD..97d3525N,
       author = {{Nakama}, Tomohiro and {Carr}, Bernard and {Silk}, Joseph},
        title = "{Limits on primordial black holes from {\ensuremath{\mu}} distortions in cosmic microwave background}",
      journal = {\prd},
         year = 2018,
        month = feb,
       volume = {97},
       number = {4},
          eid = {043525},
        pages = {043525}
}

@ARTICLE{2024ApJ...975..139Z,
       author = {{Zhang}, Saiyang and {Bromm}, Volker and {Liu}, Boyuan},
        title = "{How Do Primordial Black Holes Change the Halo Mass Function and Structure?}",
      journal = {\apj},
         year = 2024,
        month = nov,
       volume = {975},
       number = {1},
          eid = {139},
        pages = {139}
}

@ARTICLE{2010A&ARv..18....1D,
       author = {{De Zotti}, Gianfranco and {Massardi}, Marcella and {Negrello}, Mattia and {Wall}, Jasper},
        title = "{Radio and millimeter continuum surveys and their astrophysical implications}",
      journal = {\aapr},
         year = 2010,
        month = feb,
       volume = {18},
       number = {1},
        pages = {1-65}
}

@ARTICLE{2012A&A...544A..34P,
       author = {{Pushkarev}, A.~B. and {Kovalev}, Y.~Y.},
        title = "{Single-epoch VLBI imaging study of bright active galactic nuclei at 2 GHz and 8 GHz}",
      journal = {\aap},
         year = 2012,
        month = aug,
       volume = {544},
          eid = {A34},
        pages = {A34}
}

@ARTICLE{2023MNRAS.522.5434D,
       author = {{Dike}, Veronica and {Gilman}, Daniel and {Treu}, Tommaso},
        title = "{Strong lensing constraints on primordial black holes as a dark matter candidate}",
      journal = {\mnras},
         year = 2023,
        month = jul,
       volume = {522},
       number = {4},
        pages = {5434-5441}
}

@ARTICLE{2020PhRvR...2b3204S,
       author = {{Serpico}, Pasquale D. and {Poulin}, Vivian and {Inman}, Derek and {Kohri}, Kazunori},
        title = "{Cosmic microwave background bounds on primordial black holes including dark matter halo accretion}",
      journal = {Physical Review Research},
         year = 2020,
        month = may,
       volume = {2},
       number = {2},
          eid = {023204},
        pages = {023204}
}

@ARTICLE{2026PhRvL.136q1402I,
       author = {{Ivanov}, Mikhail M. and {Trifinopoulos}, Sokratis},
        title = "{Effective Field Theory Constraints on Primordial Black Holes from the High-Redshift Lyman-{\ensuremath{\alpha}} Forest}",
      journal = {\prl},
         year = 2026,
        month = may,
       volume = {136},
       number = {17},
          eid = {171402},
        pages = {171402}
}

@ARTICLE{2018MNRAS.478.3756C,
       author = {{Carr}, Bernard and {Silk}, Joseph},
        title = "{Primordial black holes as generators of cosmic structures}",
      journal = {\mnras},
         year = 2018,
        month = aug,
       volume = {478},
       number = {3},
        pages = {3756-3775}
}

@ARTICLE{2024PhRvD.110g5011G,
       author = {{Graham}, Peter W. and {Ramani}, Harikrishnan},
        title = "{Constraints on dark matter from dynamical heating of stars in ultrafaint dwarfs. I. MACHOs and primordial black holes}",
      journal = {\prd},
         year = 2024,
        month = oct,
       volume = {110},
       number = {7},
          eid = {075011},
        pages = {075011}
}

@ARTICLE{1992ARA&A..30..311B,
       author = {{Blandford}, R.~D. and {Narayan}, R.},
        title = "{Cosmological applications of gravitational lensing.}",
      journal = {\araa},
         year = 1992,
        month = jan,
       volume = {30},
        pages = {311-358}
}

@INPROCEEDINGS{2015aska.confE.143P,
       author = {{Paragi}, Z. and {Godfrey}, L. and {Reynolds}, C. and {Rioja}, M.~J. and {Deller}, A. and {Zhang}, B. and {Gurvits}, L. and {Bietenholz}, M. and {Szomoru}, A. and {Bignall}, H.~E. and {Boven}, P. and {Charlot}, P. and {Dodson}, R. and {Frey}, S. and {Garrett}, M.~A. and {Imai}, H. and {Lobanov}, A. and {Reid}, M.~J. and {Ros}, E. and {van Langevelde}, H.~J. and {Zensus}, A.~J. and {Zheng}, X.~W. and {Alberdi}, A. and {Agudo}, I. and {An}, T. and {Argo}, M. and {Beswick}, R. and {Biggs}, A. and {Brunthaler}, A. and {Campbell}, B. and {Cimo}, G. and {Colomer}, F. and {Corbel}, S. and {Conway}, J.~E. and {Cseh}, D. and {Deane}, R. and {Falcke}, H.~D.~E. and {Gawronski}, M. and {Gaylard}, M. and {Giovannini}, G. and {Giroletti}, M. and {Goddi}, C. and {Goedhart}, S. and {G{\'o}mez}, J.~L. and {Gunn}, A. and {Kharb}, P. and {Kloeckner}, H.~R. and {Koerding}, E. and {Kovalev}, Y. and {Kunert-Bajraszewska}, M. and {Lindqvist}, M. and {Lister}, M. and {Mantovani}, F. and {Marti-Vidal}, I. and {Mezcua}, M. and {McKean}, J. and {Middelberg}, E. and {Miller-Jones}, J.~C.~A. and {Moldon}, J. and {Muxlow}, T. and {O'Brien}, T. and {Perez-Torres}, M. and {Pogrebenko}, S.~V. and {Quick}, J. and {Rushton}, A. and {Schilizzi}, R. and {Smirnov}, O. and {Sohn}, B.~W. and {Surcis}, G. and {Taylor}, G.~B. and {Tingay}, S. and {Tudose}, V.~M. and {van der Horst}, A. and {van Leeuwen}, J. and {Venturi}, T. and {Vermeulen}, R. and {Vlemmings}, W.~H.~T. and {de Witt}, A. and {Wucknitz}, O. and {Yang}, J. and {Gab{\"a}nyi}, K. and {Jung}, T.},
        title = "{Very Long Baseline Interferometry with the SKA}",
    booktitle = {Advancing Astrophysics with the Square Kilometre Array (AASKA14)},
         year = 2015,
        month = apr,
          eid = {143},
        pages = {143}
}
\bibliographystyle{aasjournal}


\end{document}